\documentclass[aps,prd,preprintnumbers,nofootinbib,onecolumn,superscriptaddress]{revtex4-2}
\usepackage[colorlinks,linkcolor=blue,anchorcolor=violet,citecolor=red]{hyperref}
\usepackage{amsmath,amssymb}
\usepackage[dvipdfmx]{graphicx}
\usepackage{float}
\usepackage{comment}
\usepackage{xcolor}

\usepackage{bm}

\definecolor{darkred}{rgb}{1, 0.1, 0.3}
\definecolor{darkblue}{rgb}{0.1, 0.1, 1}
\definecolor{darkgreen}{rgb}{0,0.6,0.5}

\newcommand {\mm}[1] {\ifmmode{#1}\else{\mbox{\(#1\)}}\fi}

\newcommand{\Tr}{{\rm Tr}}

\begin{document}

\title{Leggett-Garg inequalities for testing quantumness of gravity}

\author{Akira Matsumura}
\affiliation{Department of Physics, Kyushu University, 744 Motooka, Nishi-Ku, Fukuoka 819-0395, Japan}

\author{Yasusada Nambu} 
\affiliation{Department of Physics, Nagoya University, Furo-cho, Chikusa-ku, Nagoya 464-8602, Japan}

\author{Kazuhiro Yamamoto}
\affiliation{Department of Physics,  Kyushu University, 744 Motooka, Nishi-Ku, Fukuoka 819-0395, Japan}
\affiliation{
Research Center for Advanced Particle Physics, Kyushu University, 744 Motooka, Nishi-ku, Fukuoka 819-0395, Japan}

\date{\today}

\begin{abstract}
In this study, we determine a violation of the Leggett-Garg inequalities due to gravitational interaction in a hybrid system consisting of a harmonic oscillator and a spatially localized superposed particle.
The violation of the Leggett-Garg inequalities is discussed 
using the two-time quasiprobability in connection with the entanglement negativity generated by gravitational interaction. It is demonstrated that the entanglement suppresses the violation of the Leggett-Garg inequalities when one of the two times of the quasiprobability $t_1$ is chosen as the initial time. Further, it is shown that the Leggett-Garg inequalities are generally violated due to gravitational interaction by properly choosing the configuration of the parameters, including $t_1$ and $t_2$, which are the times of the two-time quasiprobability. The feasibility of detecting violations of the Leggett-Garg inequalities in hybrid systems is also discussed.
\end{abstract}
\maketitle

\section{Introduction}
Unifying quantum mechanics and gravity is one of the most fundamental issues in physics. Feynmann discussed the possibility of testing whether gravity follows the framework of quantum mechanics \cite{Feynmann}, which has been re-examined because of recent developments in quantum information theory and quantum technologies \cite{Tabletop}.  
The proposal of testing the quantumness of gravity \cite{Bose,MV,Christodoulou}, called the BMV experiment, 
has garnered more attention and has stimulated many studies (e.g., \cite{Gnuyen,Anastopoulos,Miki} and references therein). 
The BMV experiment relies on the entanglement generated by the gravitational interaction, 
which is a quantum feature smoking gun used to characterize the nonlocal quantum interaction. 
Optomechanical systems are also promising in detecting the quantum entanglement 
generated by gravitational interaction \cite{Aspelmeyer,Schmole,Lopez, Blaushi,Miao,Matsumura,Krisnanda,Miki2}. 

Possible other approaches to detect the quantumness of gravity have been discussed in the literature. One of the 
possible approach is a non-Gaussian feature of the quantum state generated through 
the quantum force of gravity in Bose-Einstein condensate \cite{Nongaussian}. 
The authors of Ref.~\cite{Carney} have argued the visibility function of
interference in a hybrid system consisting of an oscillator and a 
particle in a spatially localized superposition state (see Fig.~\ref{config}). 
Based on Ref.~\cite{Carney}, the authors concluded that the revival in 
oscillating feature of the visibility function reflects the non-separable feature of 
the gravitational interaction, which generates the entanglement in the hybrid system
(see also \cite{Streltsov,Carney2,Maetal,Hosten}).
Therefore, it provides a unique approach to test the quantumness of gravitational interaction.

In this study, we propose a different approach to test the quantumness of gravity:
We employ the Leggett-Garg inequalities, which were proposed to 
test the macrorealism in Ref.~\cite{Leggett-Garg} (see also \cite{Emary} for a review).
Macrorealism involves characterizing classical systems, in which a macroscopic system 
is in a definite state at any given time in different available states, and the state 
can be measured without any effect on the system.
The Leggett-Garg inequalities are temporal correlations, which might be realized in a similar
analogy to the spatial nonlocal correlation described 
by Clauser-Horne-Shimony-Holt (CHSH) inequalities. 
Quantum systems may violate the predictions of macrorealism represented by the Leggett-Garg inequalities.
The violation of the Leggett-Garg inequalities has been theoretically investigated
and experimentally verified in many systems (Refs.~\cite{BHM,Saito}, and references therein). 
In this study, we apply the two-time quasiprobability introduced in Ref.~\cite{Page}, and explored
in \cite{Halliwell2016,Halliwell2019,Halliwell2021a,Halliwell2021b},
for the hybrid system described in Ref.~\cite{Carney}, to probe the quantumness of gravitational interaction. 

The remainder of this study is organized as follows. In Sec.~II, we briefly review the Leggett-Garg inequalities
based on the two-time quasiprobability and the hybrid system in Ref.~\cite{Carney}.
In Sec. III, we apply the formalism to a hybrid system, where the behavior of the 
two-time quasiprobability is examined. Feasibility of detecting the violation of the 
Leggett-Garg inequalities is also mentioned. 
In Sec. IV, the prediction within the Newton-Schr\"odinger approach is presented. 
Sec. V is a summary and conclusions.
The origin of violation of the Leggett-Garg inequalities due to gravitational interaction 
is also discussed. In the Appendix A, a deviation of Eq.~(\ref{EqSecV}) is described. 
Note that we adopt the unit $\hbar =1$ unless noted otherwise.

\section{Formulation}
\subsection{Leggett-Garg inequalities}
We begin with briefly reviewing the two-time quasiprobability
function \cite{Page,Halliwell2016,Halliwell2019,Halliwell2021a,Halliwell2021b}. 
We introduce a dichotomic variable $\hat Q=\bm n \cdot \bm\sigma$, 
where $\bm n $ is a unit vector and $\bm \sigma=(\sigma^x,\sigma^y,\sigma^z)$
is the Pauli's spin matrix. As the dichotomic variable is regarded as a spin,
$\hat Q$ is the quantum variable that gives spin value $\pm 1$ by measurement in the direction $\bm n$.
Therefore, $|\bm n\cdot \bm \sigma|^2=1$. 
The measurement operator of the dichotomic variable to obtain the measurement result $a=\pm1$ is defined as, 
\begin{eqnarray}
  \hat M_a= {1\over 2}(\bm 1+a\bm n\cdot \bm  \sigma), ~~~~~
\end{eqnarray}
which satisfies $ \hat M_a= \hat M_a^\dagger= \hat M_a^2$.

Assuming the initial state $\rho_0$, the probability that $a$ is obtained through a measurement at $t_1$ is given by
\begin{eqnarray}
  P_{1}(a)&=&\Tr\left[ \hat M_a \hat U(t_1)\rho_0 \hat U^\dagger(t_1) \hat M_a^\dagger \right]
  =\Tr[ \hat M_a(t_1)\rho_0 \hat M_a^\dagger(t_1)],
\end{eqnarray}
where we defined
\begin{eqnarray}
   \hat M_a(t)= \hat U^\dagger(t)  \hat M_a  \hat U(t)
\end{eqnarray}
and $\hat U(t)$ is the unitary operator of time evolution of the system, 
in which we assume the time-translation invariance. 
Then, the expectation value of the dichotomic variable $\hat Q$ at $t$ is 
\begin{eqnarray}
  \langle \hat Q(t)\rangle &=&\sum_{a=\pm 1} aP_1(a)
=\Tr[\bm n \cdot \bm \sigma(t_1)\rho_0],
\end{eqnarray}
where $\bm \sigma(t)=U^\dagger(t) \bm \sigma U(t)$.

Similarly, the probability that the measurement results $a$ and $b$ are obtained via 
measurements at $t_1$ and $t_2~(\geq t_1)$ with measurement axis $\bm n$
\begin{eqnarray}
  P_{12}(a,b)&=&\Tr\left[ \hat M_b \hat U(t_2-t_1) \hat M_a \hat U(t_1)\rho_0 \hat U^\dagger(t_1)M_a^\dagger  \hat U(t_2-t_1) \hat M_b\right]\nonumber\\
  &=&\Tr[ \hat M_b(t_2) \hat M_a(t_1)\rho_0 \hat M_a^\dagger(t_1) \hat M_b^\dagger(t_2)]=\Tr[ \hat M_b(t_2) \hat M_a(t_1)\rho_0 \hat M_a^\dagger(t_1)],
\end{eqnarray}
where
\begin{eqnarray}
   \hat M_b= {1\over 2}(\bm 1+b\bm n\cdot \bm \sigma).
\end{eqnarray}
The two-time correlation function is introduced as 
\begin{eqnarray}
  C(t_1,t_2)=\sum_{a,b=\pm1} ab P_{12}(a,b),
\end{eqnarray}
which reduces to 
\begin{eqnarray}
  C(t_1,t_2)={1\over 2}\Tr\left[\{\bm n\cdot\bm \sigma(t_1),\bm n\cdot\bm\sigma(t_2)\}\rho_0\right],
\end{eqnarray}
where $\{~,~\}$ denotes an anti-commutator. 

In a theory of macrorealism, the corresponding variables 
$Q_1=Q(t_1)$ and $Q_2=Q(t_2)$ take definite
values of $\pm1$, implying that 
\begin{eqnarray}
  (1+s_1Q_1)(1+s_2Q_2)\geq 0,
\end{eqnarray}
where $s_1, s_2=\pm1$. Following the framework of the macrorealism, 
there exists a joint probability distribution for the results of measurements. 
The existence of such a joint probability distribution means that we can simply average 
the above formula, and obtain the two-time Leggett-Garg inequalities \cite{Page,Halliwell2019}
\begin{eqnarray}
  1+s_1\langle Q\rangle +s_2 \langle Q_2\rangle +s_1s_2 \langle Q_1 Q_2\rangle \geq 0.
\end{eqnarray}

In the quantum mechanics, the corresponding expression can be
discussed with the two-time quasiprobability defined by 
\begin{eqnarray}
  q_{s_1,s_2}(t_1,t_2)={1\over 4}\left(1+s_1\langle \hat Q(t_1)\rangle +s_2 \langle \hat Q(t_2)\rangle +s_1s_2 C(t_1,t_2)\right),
  \label{quasiprobability}
\end{eqnarray}
which is equivalently written as \cite{Halliwell2019}
\begin{eqnarray}   &&q_{s_1s_2}(t_1,t_2)
={1\over 4}{\rm Tr}\left[{1\over2}\{1+s_1\bm n \cdot \bm \sigma(t_1),
1+s_2\bm n \cdot \bm \sigma(t_2)\}\rho_0
\right]={\rm Re}\left({\rm Tr}[M_{s_2}(t_2)M_{s_1}(t_1)\rho_0]\right).
  \label{quasiprobabilityb}
\end{eqnarray}
Note that the two-time quasiprobability produces the relations \cite{Halliwell2019}
\begin{eqnarray}
\langle \hat Q(t_1)\rangle =\sum_{a,b=\pm1} aq_{a,b}(t_1,t_2),~~~~~~ 
\langle \hat Q(t_2)\rangle =\sum_{a,b=\pm1} bq_{a,b}(t_1,t_2),~~~~~~
C_{1,2}(t_1,t_2)=\sum_{a,b=\pm1} abq_{a,b}(t_1,t_2), 
\end{eqnarray}
However, it may take negative values, which means a violation of the Leggett-Garg inequalities.

\begin{figure}[t]
\centering
\includegraphics[width=5.0in]{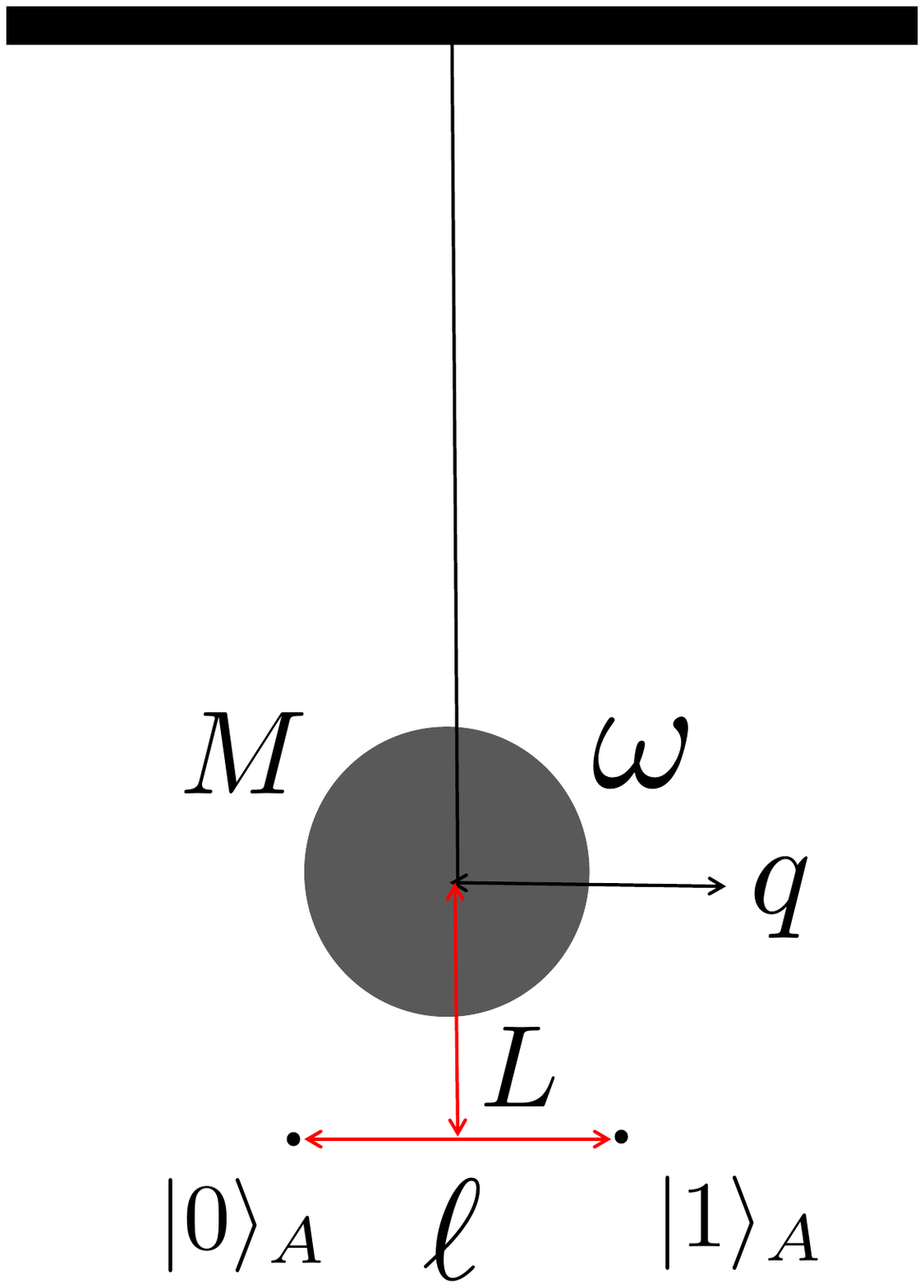}
\caption{System consisting of an oscillator and a particle. The particle is in a superposition state of
  two spatially localized states denoted by $|0\rangle_A$ and $|1\rangle_A$.
  The position of the oscillator is denoted by $q$, and its mass and angular frequencies are
  $M$ and $\omega$, respectively.
$L$ is the distance between the oscillator and the particle in a superposition state, 
and $\ell$ is the distance between the positions of the two spatially localized states. 
  \label{config}}
\end{figure}
\subsection{Hybrid system}
We consider a hybrid system consisting of an oscillator and a particle (see Figure \ref{config}). An oscillator with a mass $M$ is described by the coordinate variable $q$, whose oscillation is characterized by the angular frequency $\omega$. A particle with mass $m$ is in a superposition of the two spatially localized states denoted by $|0\rangle_A$ and $|1\rangle_A$. 
Here, we assume that $\ell$ is the distance between the positions of the two spatially localized states, and $L$ is the distance between the oscillator and the particle. 
This model was introduced in Ref. \cite{Carney}, and the authors investigated the effects of gravitational 
interaction between the oscillator and the particle on the visibility function, owing to the interference of the particle's state.
Earlier report \cite{Carney} demonstrated that a revival of a visibility function owing to the interference
is the result of the entanglement because of the gravitational interaction, 
which can be tested as a signature of the quantumness of gravity. 
Furthermore, the non-separable evolution owing to gravitational interaction is more fundamental for their argument to generate entanglement \cite{Carney,Carney2}.

We investigate the Leggett-Garg inequalities in a hybrid system, whose 
Hamiltonian is given by 
\begin{eqnarray}
H=\Omega\sigma^z  +\omega a^\dagger a + H_{\rm grav},
\label{HamiltonianCarny}
  \end{eqnarray}
where $\omega a^\dagger a$ is a free Hamiltonian of the oscillator
with the creation (annihilation) operator $a ~(a^\dagger)$,
and the last term of the right-hand side of Eq.~(\ref{HamiltonianCarny}) describes the
gravitational interaction between the oscillator and the particle.
The eigenstates of $\sigma^z$ describe the two spatially localized states of the particle, 
and the first term of Eq.~(\ref{HamiltonianCarny}), $\Omega\sigma^z$, causes 
the phenomenon corresponding to the Larmor precession in the two states,
which is not included in the analysis of Ref. \cite{Carney}.
Following the configuration as shown in Fig.\ref{config}, the gravitational potential 
of the system can be written as,
\begin{eqnarray}
  H_{\rm grav}=-{GMm\over \sqrt{L^2+(q+\sigma^z \ell /2)^2}}\simeq {GMm q\ell \sigma^z\over \sqrt{L^2+\ell^2/4}^3}+{\rm constant},
\end{eqnarray}
where $G$ is the Newton constant and the approximate expression is obtained by 
assuming that $q$ is small compared to $L$ and $\ell$. Introducing constant $g$ and nondimensional 
variable $\tilde q$ by 
\begin{eqnarray}
  g={GMm \ell \over \sqrt{L^2+\ell^2/4}^3}{1\over \sqrt{2M\omega}}, ~~~
  q={1\over \sqrt{2M\omega}}\sqrt{2}\tilde q,
\end{eqnarray}
the Hamiltonian of the gravitational interaction reduces to
\begin{eqnarray}
  H_{\rm grav}=g\sigma^z\sqrt{2}\tilde q.
\end{eqnarray}

The unitary operator of the Hamiltonian is written as
\begin{eqnarray}
  U(t)=e^{-iHt}=e^{-i(\Omega\sigma^z  +\omega a^\dagger a)t}
  T\exp\left[-i\int_0^t dt'g\sigma^z\sqrt{2}\tilde q_I(t)\right],
\end{eqnarray}
where $\tilde q_I$ denotes $\tilde q$ in the interaction picture,
\begin{eqnarray}
  \tilde q_I(t)=e^{i\omega a^\dagger at}\tilde qe^{-i\omega a^\dagger at}=
{1\over \sqrt{2}}(e^{-i\omega t}a+e^{i\omega t}a^\dagger).
\end{eqnarray} 
Note that $\sigma^z$ in the interaction picture is
$\sigma_{I}^z(t)=e^{i\Omega\sigma^zt}\sigma^ze^{-i\Omega\sigma^zt}=\sigma^z$. 
Using the following relation (see also \cite{Matsumura}), we have
\begin{eqnarray}
  T\exp\left[-i\int_0^t dt'g\sigma^z\sqrt{2}\tilde q_I(t')\right]
  &=&\exp\left[
    -i\int_0^t dt'g\sigma^z
    \sqrt{2}\tilde q_I(t')
    -g^2 \int_0^{t}dt'\int_0^{t'}dt''[\tilde q_I(t'),\tilde q_I(t'')]
    \right]
    \nonumber\\
    &=&e^{
    g\sigma^z(\alpha(t)a-\alpha^*(t)a^\dagger)
    +ig^2 \beta(t)
    },
\end{eqnarray}We used the following relations to derive the second equality
\begin{eqnarray}
  &&-i\int_0^t \sqrt{2}\tilde q_I(t')
  dt'=\alpha(t)a-\alpha^*(t)a^\dagger
  \\
 &&\int_0^{t}dt'\int_0^{t'}dt'' [\tilde q_I(t'),\tilde q_I(t'')]=-i\beta(t),
\end{eqnarray}
and defined
\begin{eqnarray}
  &&\alpha(t)={e^{-i\omega t}-1\over \omega},~~~~~~
  \beta(t)={1\over \omega}\left(t-{\sin\omega t\over \omega}\right).
\end{eqnarray}
Excepting the total phase, the unitary operator of time evolution of the system
is written as
\begin{eqnarray}
  U(t)=e^{-i(\Omega\sigma^z+\omega  a^\dagger a)t} e^{g\sigma^z(\alpha(t)a-\alpha^*(t)a^\dagger)}.
  \label{Unitary}
\end{eqnarray}

\subsection{Two-time Quasiprobability}
We determine the two-time quasiprobability for the particle in the hybrid system above when
the initial state is prepared as 
\begin{eqnarray}
|\psi_0\rangle={1\over \sqrt{2}}(|0\rangle_A+|1\rangle_A)\otimes|0\rangle,
\label{isvacuum}
\end{eqnarray}
where $|0\rangle$  is the ground state of the oscillator.
Using the unitary operator (\ref{Unitary}), the state at time $t$ is: 
\begin{eqnarray}
|\psi(t)\rangle =U(t)|\psi_0\rangle
&=& {e^{-i\omega a^\dagger at}\over \sqrt{2}}\left(e^{-i\Omega t}|0\rangle_A |-g\alpha^*(t)\rangle_C+
e^{i\Omega t}|1\rangle_A |+g\alpha^*(t)\rangle_C\right),
\nonumber\\
&=&{1\over \sqrt{2}}\left(e^{-i\Omega t}|0\rangle_A |g\alpha(t)\rangle_C+
e^{i\Omega t}|1\rangle_A |-g\alpha(t)\rangle_C\right),
\label{ehqsc}
\end{eqnarray}
where the oscillation is in the coherent state $|\xi\rangle_C$  defined by
$|\xi\rangle_C = e^{\xi a^\dagger-\xi^* a}|0\rangle$.
In deriving the second line of the equation, we used the expression of the
coherent states in the Fock basis,
\begin{eqnarray}
|\xi\rangle_C=e^{-|\xi|^2/2}\sum_{m=0}^\infty{\xi^m\over \sqrt{m!}}|m\rangle
\nonumber
\end{eqnarray}
where $|m\rangle$ is the $m$th energy excited state of the oscillator.

Now, we determine the expression of the two-time quasiprobability function (\ref{quasiprobability}).
Hereafter we consider the case 
\begin{eqnarray}
  \bm n=(\cos \varphi, \sin\varphi, 0), ~~~~
\end{eqnarray}
unless otherwise stated.
For the initial state $\rho_0=|\psi_0\rangle\langle\psi_0|$ with (\ref{isvacuum}), from straightforward computations, we obtain
\begin{eqnarray}
  &&\langle \hat Q_1 \rangle =\Tr[\bm n\cdot\bm\sigma(t_1) \rho_0]=
  \cos(2\Omega t_1-\varphi) e^{-8{\lambda^2}{\sin^2{\omega t_1\over 2}}}
  \\
  &&\langle \hat Q_2 \rangle =\Tr[\bm n\cdot\bm\sigma(t_2) \rho_0]=
    \cos(2\Omega t_2-\varphi) e^{-8{\lambda^2}{\sin^2{\omega t_2\over 2}}}
\end{eqnarray}
and
\begin{eqnarray}
  C(t_2,t_1)={1\over 2}\Tr\left[\{\bm n\cdot\bm \sigma(t_1),\bm n\cdot\bm\sigma(t_2)\}\rho_0\right] =\cos\Theta(t_2,t_1)\cos(2\Omega(t_2-t_1))e^{-8{\lambda^2}{\sin^2{\omega (t_2-t_1)\over 2}}},
\end{eqnarray}
where we defined
\begin{eqnarray}
\Theta(t_2,t_1)=4\lambda^2(\sin\omega(t_2-t_1)
-\sin\omega t_2+\sin\omega t_1)=16\lambda^2\sin{\omega(t_2-t_1)\over 2}
\sin{\omega t_2\over 2}\sin{\omega t_1\over 2} 
\label{DefTheta}
\end{eqnarray}
and
\begin{eqnarray}
\lambda={g\over \omega}.
\end{eqnarray}
Then, the expression for the two-time quasiprobability is written as:
\begin{eqnarray}
  &&q_{s_1s_2}(t_1,t_2)={1\over 4}\Bigl(1+s_1\cos(2\Omega t_1-\varphi) e^{-8{\lambda^2}\sin^2{\omega t_1\over 2}}
  +s_2\cos(2\Omega t_2-\varphi) e^{-8{\lambda^2}{\sin^2{\omega t_2\over 2}}}
  \nonumber\\
  &&~~~~~~~~~~~~~~~~~~~~~
  +s_1s_2\cos\Theta(t_2,t_1)\cos(2\Omega(t_2-t_1))e^{-8{\lambda^2}\sin^2{\omega (t_2-t_1)\over 2}}\Bigr).
  \label{ttqp}
\end{eqnarray}

\section{Behavior of two-time quasiprobability}
\subsection{Case of  $t_1=0$ and $\Omega\neq0$}
In this section, we investigate the behavior of the two-time quasiprobability. 
We first consider the cases imposing that $t_1$ is the initial time, $t_1=0$,
and $\Omega\neq0$
in Eq.~(\ref{ttqp}).
In this case, we show that gravitational interaction suppresses the violation of the
Leggett-Garg inequalities. Imposing $t_1=0$ on the two-time quasiprobability (\ref{ttqp}),
we have
\begin{eqnarray}
  q_{s_1,s_2}(0,t_2)
  &=&
  {1\over 4}\Bigl(1+s_1\cos\varphi
  +s_2\cos(2\Omega t_2-\varphi) e^{-8{\lambda^2}{\sin^2{\omega t_2\over 2}}}
  +s_1s_2\cos(2\Omega t_2)e^{-8{\lambda^2}\sin^2{\omega t_2\over 2}}\Bigr).
\end{eqnarray}
Assuming that the Leggett-Garg inequalities is violated when the gravitational interaction 
is switched off by setting $\lambda=0$ 
\begin{eqnarray}
 1+s_1\cos\varphi
  +s_2\cos(2\Omega t_2-\varphi) 
  +s_1s_2\cos(2\Omega t_2)<0.
\end{eqnarray}
This inequality holds, depending on the parameters excepting $\varphi=0$.
Under this condition, we have
\begin{eqnarray}
  s_2\cos(2\Omega t_2-\varphi) 
  +s_1s_2\cos(2\Omega t_2)<0
  \label{because}
\end{eqnarray}
because $1+s_1\cos\varphi\geq0$ is always satisfied. 
Then, the quasiprobability is rewritten as
\begin{eqnarray}
  q_{s_1,s_2}(0,t_2)
  &=&{1\over 4}\Bigl(1+s_1\cos\varphi 
  +s_2\cos(2\Omega t_2-\varphi)
  +s_1s_2\cos(2\Omega t_2)\Bigr)
  \nonumber
    \\
  &&-{1\over 4}(1-e^{-8{\lambda^2}\sin^2{\omega t_2\over 2}} )\left(s_2\cos(2\Omega t_2-\varphi) 
  +s_1s_2\cos(2\Omega t_2)\right).
\label{qs1s20t2}
\end{eqnarray}
The terms in the second line of Eq (\ref{qs1s20t2}),
which originates from gravitational interaction, are always positive from Eq (\ref{because}).
This means that gravitational interaction always suppresses the violation of the Leggett-Garg inequalities in this case. 

Because the
gravitational interaction generates the entanglement between the oscillator and the
particle, the above argument is rephrased using the entanglement. 
To quantify the entanglement of a given density matrix 
$\rho_{12}$ of a bipartite system, we use the entanglement negativity \cite{Vidal}, 
\begin{eqnarray}
N=\sum_{\lambda_i <0} |\lambda_i|,
\label{N}
\end{eqnarray}
where 
$\lambda_i$ is the eigenvalue of the partial transpose 
$\rho^{T_1}_{12}$ with the elements ${}_1 \langle i| {}_2 \langle j| \rho^{T_1}_{12} |k \rangle_1 |\ell \rangle_2={}_1 \langle k| {}_2 \langle j| \rho_{12} |i \rangle_1 |\ell \rangle_2$. 
The evolved state 
$|\psi(t) \rangle$ is rewritten as 
\begin{eqnarray}
|\psi(t) \rangle
&=&
{1\over \sqrt{2}}\left(e^{-i\Omega t}|0\rangle_A |g\alpha(t)\rangle_C+
e^{i\Omega t}|1\rangle_A |-g\alpha(t)\rangle_C\right)
\nonumber 
\\
&=&{1\over \sqrt{2}}
\left[e^{-i\Omega t}|0\rangle_A 
\left(
{\sqrt{N_+}\over 2}|+\rangle_C+{\sqrt{N_-}\over 2}|-\rangle_C
\right)
+
e^{i\Omega t}|1\rangle_A \left(
{\sqrt{N_+}\over 2}|+\rangle_C
-{\sqrt{N_-}\over 2}|-\rangle_C
\right)
\right],
\end{eqnarray}
where 
$|\pm \rangle_C=1/\sqrt{N_\pm}(|g\alpha(t)\rangle_C \pm |-g\alpha(t)\rangle_C)$ and
$N_\pm=2\pm 2e^{-2g^2|\alpha(t)|^2}$. 
Hence 
$|\psi(t) \rangle$ is regarded as a two-qubit state with the basis 
$\{|0\rangle_A |+\rangle_C, |0\rangle_A |-\rangle_C, |1\rangle_A|+\rangle_C, |1\rangle_A |-\rangle_C \}$
and the density matrix 
$\rho(t)=|\psi(t) \rangle \langle \psi(t)|$ is a $4\times4$ matrix.
This is due to the fact that the Schmidt rank of a pure hybrid state is always finite. 
From the partial transposed matrix 
$\rho^{T_A}(t)$, 
we obtain 
the following entanglement 
negativity
\begin{eqnarray*}
N(t)={1\over 2}\sqrt{1-e^{-16\lambda^2\sin^2{\omega t\over 2}}}.
\end{eqnarray*}
In e.g., \cite{Arkhipov}, the above procedure was performed 
for a pure hybrid qubit-Schr\"{o}dinger cat state.

The term of the gravitational interaction in \eqref{qs1s20t2} is expressed as 
\begin{eqnarray}
  1-e^{-8\lambda^2 \sin^2{\omega t\over 2}}=1-\sqrt{1-4N^2(t)}.
\end{eqnarray}
Then, Eq.~(\ref{qs1s20t2}) can be written as
\begin{eqnarray}
  q_{s_1,s_2}(0,t_2)
  &=&{1\over 4}\Bigl(1+s_1\cos\varphi 
  +s_2\cos(2\Omega t_2-\varphi)
  +s_1s_2\cos(2\Omega t_2)\Bigr)
  \nonumber
    \\
  &&-{1\over 4}(1-\sqrt{1-4N^2(t_2)})\left(s_2\cos(2\Omega t_2-\varphi) 
  +s_1s_2\cos(2\Omega t_2)\right).
\label{qs1s20t3}
\end{eqnarray}
The negativity takes values $0\leq N(t)\leq 1/2$, 
in which $1-\sqrt{1-4N^2(t)}$ is the monotonic
increasing function of $N(t)$. Therefore, 
this implies that the entanglement suppresses the
violation of the Leggett-Garg inequalities of $ q_{s_1,s_2}(0,t_2)$. 

For the $\Omega=0$ case, we can determine the relation between quasiprobability function and the negativity in the limit of $\lambda\ll1$. In this limit, we have
\begin{eqnarray}
N(t)\simeq 2\lambda \Bigl|\sin{\omega t\over 2}\Bigr|\ll1,
\end{eqnarray}
with which Eq.~(\ref{qs1s20t3}) reduces to
\begin{eqnarray}
  q_{s_1s_2}(0,t_2)&\simeq&{1\over 4}\Bigl(1+s_1\cos\varphi
  +s_2\cos\varphi+ s_1 s_2 
-2N^2(t)
  (s_2\cos\varphi+s_1 s_2)\Bigr)
\end{eqnarray}
for $\Omega=0$. Furthermore, for $s_1\cos\varphi=1$, $s_1=-s_2$, 
we have
\begin{eqnarray}
  q_{s_1s_2}(0,t)
  \simeq N^2(t).
\end{eqnarray}
Thus, by choosing suitable parameters, the quasiprobability 
reflects the evolution of the entanglement negativity directly.

\begin{figure}[!b]
\centering\includegraphics[width=2.2in]{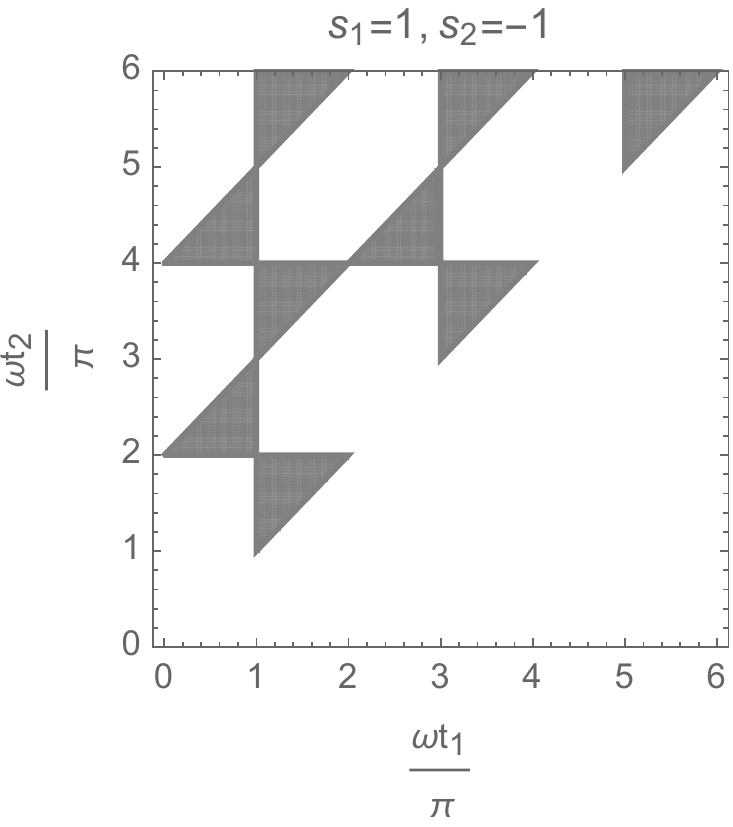}~~~~
\centering\includegraphics[width=2.2in]{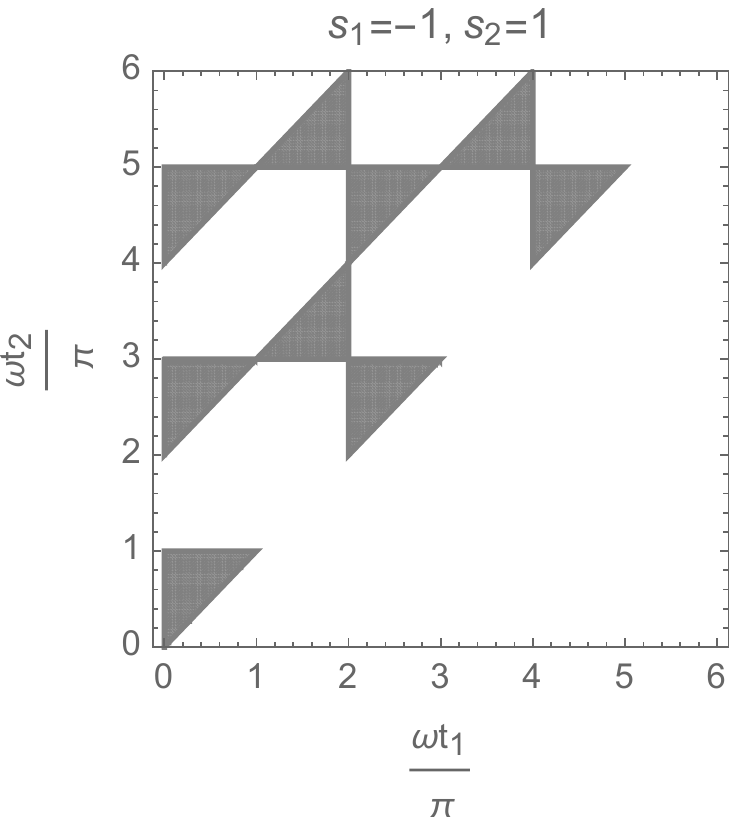}~~~~.
\centering\includegraphics[width=2.2in]{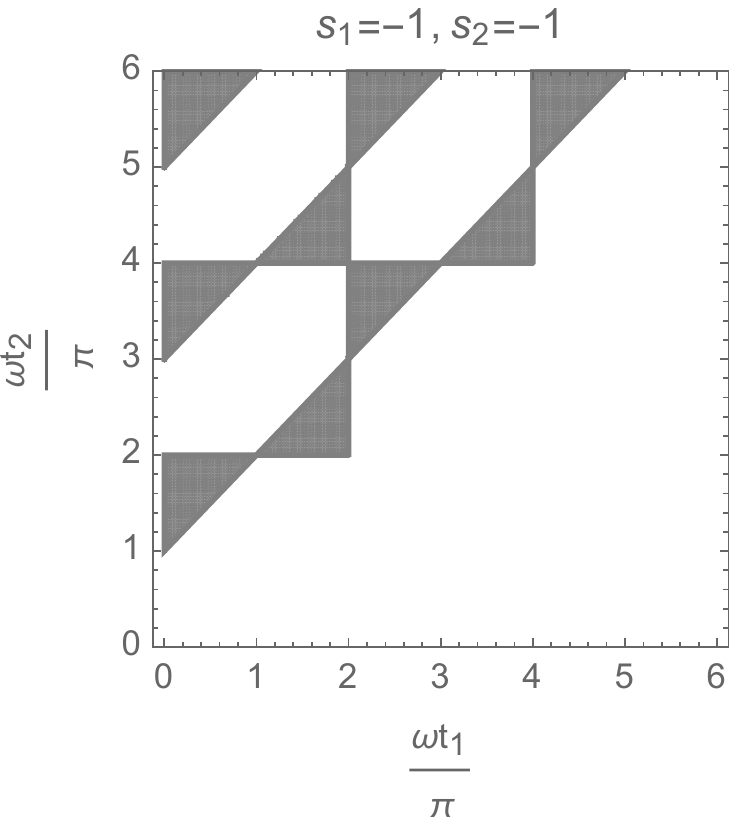}
\caption{Shaded region shows the regions where $q_{s_1s_2}(t_1,t_2)<0$ is satisfied 
on the $\omega t_1/\pi$ (horizontal axis) and $\omega t_2/\pi$ (vertical axis) planes, respectively. We adopted  $\Omega=0$, $\varphi=0$, 
$s_1=1,~s_2=-1$ (left panel), $s_1=-1,~s_2=1$ (middle panel), 
and $s_1=-1,~s_2=-1$ (right panel), 
respectively. Here, we only show the region satisfying $0\leq t_1\leq t_2$
and that we adopted $8\lambda^2=10^{-2}$.
  \label{Gmacroc}}
\end{figure}

\subsection{Case of $t_1\neq 0$ and $\Omega=0$}
Next, we consider the violation of the Leggett-Garg inequality due to the gravitational interaction by setting $\Omega=0$.
Here we assume the case where $\varphi=0$, for simplicity.
Then, the two-time quasiprobability becomes
\begin{eqnarray}
  q_{s_1s_2}(t_1,t_2)={1\over 4}\Bigl(1+s_1 e^{-8{\lambda^2}\sin^2{\omega t_1\over 2}}
  +s_2 e^{-8{\lambda^2}{\sin^2{\omega t_2\over 2}}}
  +s_1s_2\cos\Theta(t_2,t_1)
  e^{-8{\lambda^2}\sin^2{\omega (t_2-t_1)\over 2}}\Bigr).
  \label{demo}
\end{eqnarray}
For $\lambda=0$, the two-time quasiprobability satisfies
$q_{s_1s_2}(t_1,t_2)={1\over 4}(1+s_1+s_2+s_1s_2)\geq0$.
Figure \ref{Gmacroc} demonstrates the region where the quasiprobability function (\ref{demo}) with $\lambda\neq0$ takes negative values on the $t_1$ and $t_2$ planes, in which we showed the region
satisfying $0\leq t_1\leq t_2$. Thus, the Leggett-Garg inequalities are violated because of the gravitational interaction. 

When $\lambda\ll1$, the contribution from the term $\cos\Theta(t_2,t_1)$
in (\ref{demo}) becomes the highest order of ${\cal O}(\lambda^4)$. 
Then, up to the order of ${\cal O}(\lambda^2)$,
the quasiprobability (\ref{demo}) reduces to
\begin{eqnarray}
  &&q_{s_1s_2}(t_1,t_2)\simeq{1\over 4}(1+s_1 +s_2+s_1s_2)
  -2{\lambda^2}\left(s_1\sin^2{\omega t_1\over 2}+s_2\sin^2{\omega t_2\over 2}
  +s_1 s_2\sin^2{\omega (t_2-t_1)\over 2}\right),
  \label{demo2}
\end{eqnarray}
which may take negative values when $(s_1,s_2)=(1,-1)$, or $(-1,1)$, 
or $(-1,-1)$ owing to the gravitational interaction. The minimum value of the quasiprobability function is
approximately 
\begin{eqnarray}
  {\rm min} \{q_{s_1s_2}(t_1,t_2)\} \simeq -{\lambda^2\over 2},
\end{eqnarray}
which appears for $s_1=1, s_2=-1$ when
\begin{eqnarray}
&&\omega t_1={2\over 3}\pi+2\pi n, ~~\omega t_2={7\over 3}\pi +2\pi n, ~~(n=0, 1, 2,\cdots),\\
&&\omega t_1={4\over 3}\pi+2\pi m, ~~\omega t_2={5\over 3}\pi +2\pi m, ~~(m=0, 1, 2,\cdots),
\end{eqnarray}
and for $s_1=-1, s_2=1$ when 
\begin{eqnarray}
&&\omega t_1={\pi\over 3}+2\pi n, ~~\omega t_2={2\over 3}\pi +2\pi n, ~~(n=0, 1, 2,\cdots),\\
&&\omega t_1={5\over 3}\pi+2\pi m, ~~\omega t_2={10\over 3}\pi +2\pi m, ~~(m=0, 1, 2,\cdots),
\end{eqnarray}
and for $s_1=s_2=-1$ when
\begin{eqnarray}
&&\omega t_1={\pi\over 3}+2\pi n, ~~\omega t_2={5\over 3}\pi +2\pi n, ~~(n=0, 1, 2,\cdots),\\
&&\omega t_1={5\over 3}\pi+2\pi m, ~~\omega t_2={7\over 3}\pi +2\pi m, ~~(m=0, 1, 2,\cdots).
\end{eqnarray}

Summarizing the result of the case, $\Omega=0$, 
the gravitational interaction is the unique interaction to evolve 
the particle's state. In this case, the Leggett-Garg inequalities
are violated, except in the case $s_1=s_2=1$. 
The violation of the Leggett-Garg inequalities depends on the parameters, 
$s_1$, $s_2$, $t_1$, $t_2$, and $\varphi$, 
which is not explicitly shown. The minimum value of the two-time quasiprobability
is $-\lambda^2/2$. The violation further depends on the initial state, for which 
we adopted Eq.~(\ref{isvacuum}) in this subsection.
Notably, in the case $\Omega=0$, the violation of the 
Leggett-Garg inequalities is derived from the gravitational interaction 
and that there appears no violation of the Leggett-Garg inequalities 
in the absence of the gravitational interaction.

\subsection{Thermal state as initial state for oscillator}
In this subsection, we consider the effects of the initial condition on the 
Leggett-Garg inequalities. Here, we adopt a thermal state for the  initial state of the
oscillator.  The thermal state can be described by the density matrix in the Glauber P-representation on the 
basis of the coherent state
\begin{eqnarray}
  && \rho_{\rm th}={1\over \pi\bar n }\int d^2\gamma e^{-|\gamma|^2/{\bar n}}|\gamma\rangle_C{}_C\langle\gamma|,
\end{eqnarray}
where $\bar n$ is the mean occupation number, which is related to temperature $T$ by 
$\bar n=k_BT/2\omega$ with the Boltzmann constant $k_B$, 
and  $|\gamma\rangle_C$ represents the coherent state. 
Using the following expectation value with respect to the thermal state 
\begin{eqnarray}
  &&\hspace{-2cm}
  {\rm Tr}\Bigl[\rho_{\rm th} e^{\pm 2g\bigl((\alpha(t_2)-\alpha(t_1))a-(\alpha^*(t_2)-\alpha^*(t_1))a^\dagger\bigr)}\Bigr]
     \nonumber\\
     &=&{1\over  \pi\bar n }\int d^2\gamma e^{-|\gamma|^2/{\bar n}}
   {}_C\langle\gamma|\exp[\pm 2g\bigl((\alpha(t_2)-\alpha(t_1))a-(\alpha^*(t_2)-\alpha^*(t_1))a^\dagger\bigr)]|\gamma\rangle_C
   \nonumber\\
     &=&{1\over  \pi\bar n }\int d^2\gamma e^{-|\gamma|^2/{\bar n}}
 \exp\left(-2g^2
 |(\alpha(t_2)-\alpha(t_1))|^2  \right)e^{\pm 4ig{\rm Im}[(\alpha(t_2)-\alpha(t_1))\gamma]}
    \nonumber\\
     &=& \exp\left(-2(2\bar n+1)g^2
    |(\alpha(t_2)-\alpha(t_1))|^2  \right),
\end{eqnarray}
we  find
\begin{eqnarray}
  &&\langle \hat Q_1 \rangle =\Tr[\bm n\cdot\bm\sigma(t_1) \rho_{\rm th}]=
  \cos(2\Omega t_1-\varphi) e^{-8(2\bar n+1){\lambda^2}{\sin^2{\omega t_1\over 2}}}
  \\
  &&\langle \hat Q_2 \rangle =\Tr[\bm n\cdot\bm\sigma(t_2) \rho_{\rm  th}]=
    \cos(2\Omega t_2-\varphi) e^{-8(2\bar n+1){\lambda^2}{\sin^2{\omega t_2\over 2}}}
\end{eqnarray}
and 
\begin{eqnarray}
  C(t_2,t_1)
  &=&{1\over 2}\Tr\left[\{\bm n\cdot\bm \sigma(t_1),\bm n\cdot\bm\sigma(t_2)\}\rho_{\rm th}\right]
    \nonumber\\
  &=&
  \cos\Theta(t_2,t_1)\cos(2\Omega (t_2-t_1))
\exp\left(-8(2\bar n+1){\lambda^2}\sin^2{\omega(t_2-t_1)\over2}  \right)
\end{eqnarray}
with $\Theta(t_2,t_1)$ is defined by the Eq.~(\ref{DefTheta}).
Thus, the quasiprobability with the thermal state as the oscillator's initial condition is given by
\begin{eqnarray}
  &&q_{s_1s_2}(t_1,t_2)={1\over 4}\Bigl(1+s_1\cos(2\Omega t_1-\varphi) e^{-8(2\bar n+1){\lambda^2}\sin^2{\omega t_1\over 2}}
  +s_2\cos(2\Omega t_2-\varphi) e^{-8(2\bar n+1){\lambda^2}{\sin^2{\omega t_2\over 2}}}
  \nonumber\\
  &&~~~~~~~~~~~~~~~~~~~~~
  +s_1s_2\cos\Theta(t_2,t_1)\cos(2\Omega (t_2-t_1))e^{-8(2\bar n+1){\lambda^2}\sin^2{\omega (t_2-t_1)\over 2}}\Bigr).
\end{eqnarray}
The difference between the ground state and thermal state is the factor $(2\bar n+1)$
in the exponential function. 
Therefore, if $\lambda$ is small, $\lambda\ll1$, the minimum value of 
the quasiprobability function appears 
under the same condition, as the ground state of oscillator 
in the previous section with $\Omega=0$,
and the minimum value is approximately given by
\begin{eqnarray}
  {\rm min} \{q_{s_1s_2}(t_1,t_2)\} \simeq -{\lambda^2\over 2}(2\bar n+1).
\end{eqnarray}

\subsection{Squeezed state as the initial state of the oscillator}
\begin{figure}[!t]
\centering\includegraphics[width=2.2in]{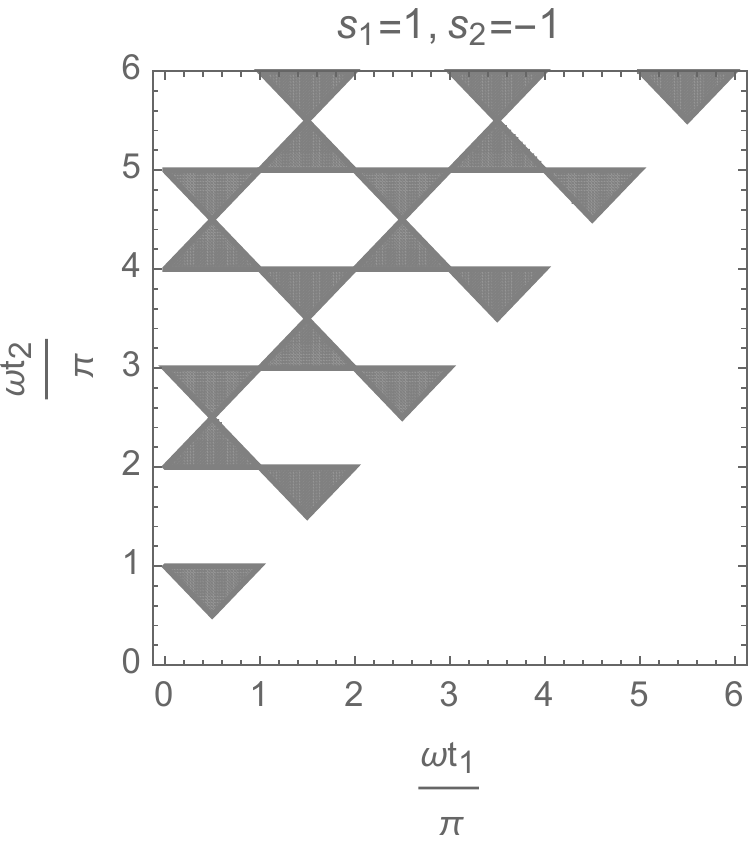}~~~~~~
\centering\includegraphics[width=2.2in]{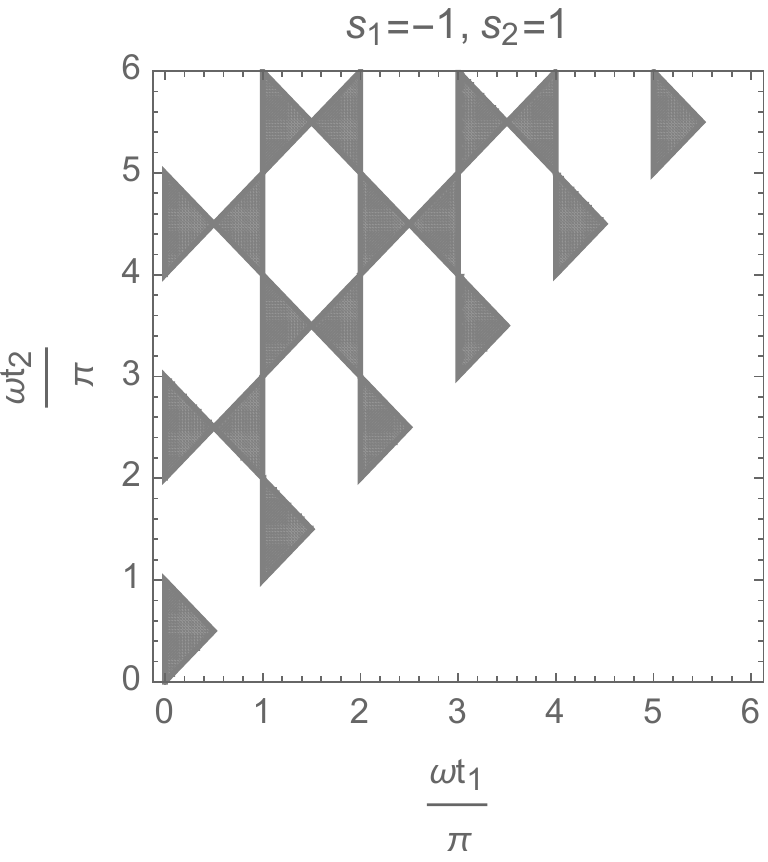}~~~~~~
\centering\includegraphics[width=2.2in]{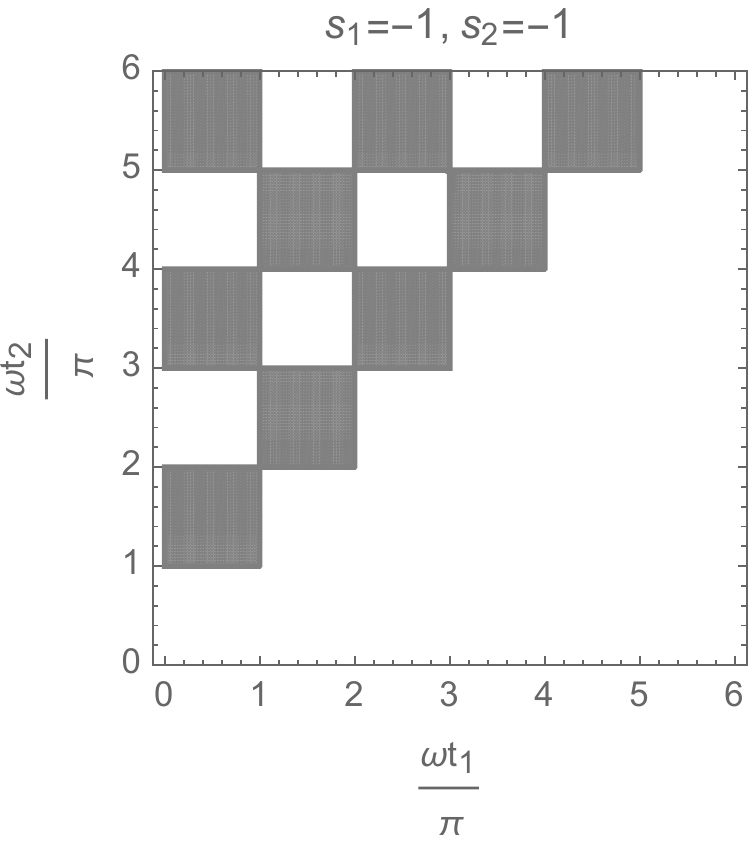}
\caption{Shaded region shows the regions satisfying $q_{s_1s_2}(t_1,t_2)<0$ 
of Eq.(\ref{qssttsq2}) with the squeezed state as the oscillator's initial condition 
on the $\omega t_1/\pi$ and $\omega t_2/\pi$ planes, respectively. We adopted  $\Omega=0$, $\varphi=0$, 
$s_1=1,~s_2=-1$ (left panel); $s_1=-1,~s_2=1$ (middle panel); 
and $s_1=-1,~s_2=-1$ (right panel), 
respectively. Here we adopted $8\lambda^2=10^{-4}$ and $\zeta=5$.
Here, we only show the region $0\leq t_1\leq t_2$.
 In this case, no violation of the Leggett-Garg inequalities appears for $s_1=s_2=1$. 
  \label{Squeeze}}
\vspace{1cm}
\centering\includegraphics[width=2.2in]{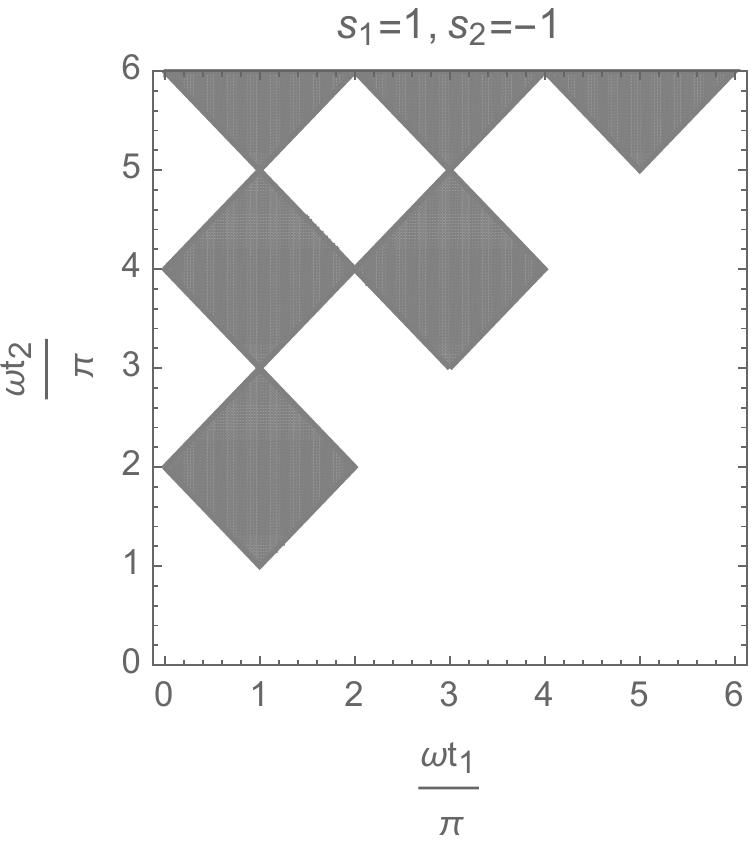}~~~~~~
\centering\includegraphics[width=2.2in]{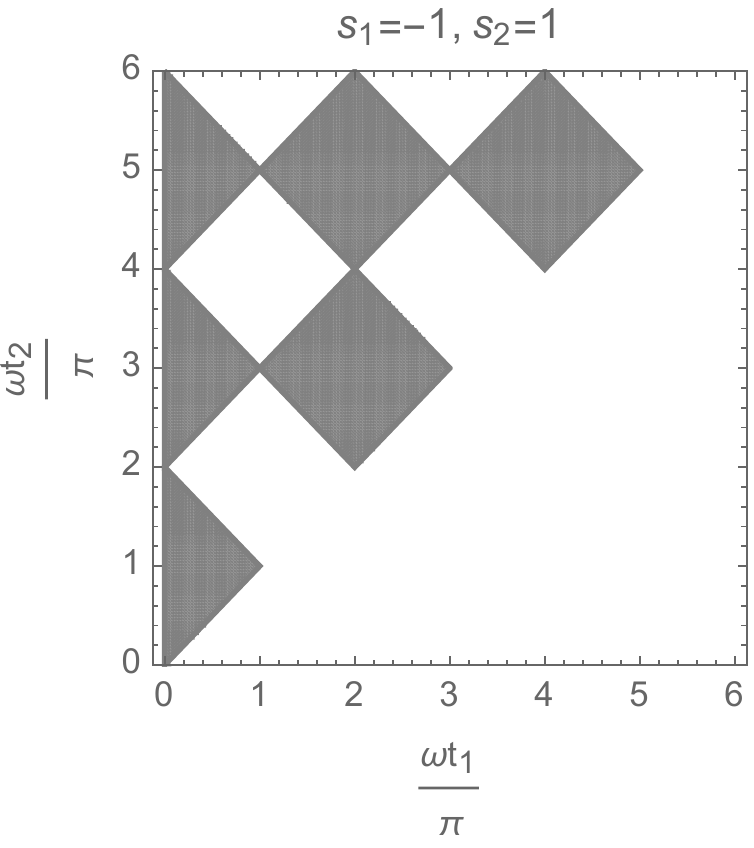}~~~~~~
\caption{Same as the panels in Fig.~\ref{Squeeze}, but with $\zeta=-5,~s_1=1,~s_2=-1$ (left panel) and
   $\zeta=-5$,~$s_1=-1,~s_2=1$ (right panel);.
 The other parameters are  $\Omega=0$, ~$\varphi=0$, and ~$8\lambda^2=10^{-4}$.
 In this case, no violation of the Leggett-Garg inequalities appears for $s_1=s_2=1$ and $s_1=s_2=-1$. 
  \label{Squeeze2}}
\end{figure}

Further, we consider the squeezed state as the initial state of the oscillator. The  squeezed state can be 
obtained by
\begin{eqnarray}
  |\zeta\rangle_S=S(\zeta)|0\rangle
\end{eqnarray}
with the squeezing operator $S(\zeta)$ defined by
$  S(\zeta)=e^{{1\over 2}(\zeta a^\dagger{}^2-\zeta^*a^2)}. 
$  
By using the mathematical formula
$ D(\xi)S(\zeta)=S(\zeta)D(\gamma)
$ 
where
$ \gamma=\xi\cosh |\zeta|-\xi^*e^{i\theta}\sinh |\zeta|$
with 
$ \zeta=|\zeta|e^{i\theta},
$ 
we determine the expectation values with the squeezed state as the initial state for the oscillator, 
\begin{eqnarray}
  \rho_{sq}=|\psi_{sq}\rangle\langle\psi_{sq}|,
  ~~~~
  |\psi_{sq}\rangle={1\over \sqrt{2}}(|0\rangle_A+|1\rangle_A)|\zeta\rangle_S,
\end{eqnarray}
as
\begin{eqnarray}
  &&
  \hspace{-1cm}\langle \hat Q_1 \rangle =\Tr[\bm n\cdot{\bm \sigma}(t_1) \rho_{\rm sq}]=
  \cos(2\Omega t_1-\varphi) \exp\left[-2{\lambda^2}
  \big|(e^{i\omega t_1}-1)\cosh |\zeta|-(e^{-i\omega t_1}-1)e^{i\theta}\sinh |\zeta|\big|^2\right]
  \\
  &&\hspace{-1cm}\langle \hat Q_2 \rangle =\Tr[\bm n\cdot\bm \sigma(t_2) \rho_{\rm  sq}]=
    \cos(2\Omega t_2-\varphi)  \exp\left[-2{\lambda^2}
  \big|(e^{i\omega t_2}-1)\cosh |\zeta|-(e^{-i\omega t_2}-1)e^{i\theta}\sinh |\zeta|\big|^2\right]
\end{eqnarray}
and
\begin{eqnarray}
  C(t_2,t_1) &=&{1\over 2}\Tr\left[\{\bm n\cdot\bm \sigma(t_1),\bm n\cdot\bm\sigma(t_2)\}\rho_{\rm sq}\right]
    \nonumber\\
  &=&\cos\Theta(t_2,t_1)\cos(2\Omega (t_2-t_1))
\exp\left[-2\lambda^2\big|(e^{i\omega t_2}-e^{i\omega t_1})\cosh |\zeta|-(e^{-i\omega t_2}-e^{-i\omega t_1})e^{i\theta}\sinh |\zeta|\big|^2)\right]    \nonumber\\
\end{eqnarray}
with $\Theta(t_2,t_1)$ is defined by the Eq.~(\ref{DefTheta}).
In $\Omega=0$ and $\varphi=0$ limits, the two-time quasiprobability reads
\begin{eqnarray}
  &&q_{s_1s_2}(t_1,t_2)={1\over 4}\biggl(1
  +s_1
  \exp\left[-2{\lambda^2}
  \big|(e^{i\omega t_1}-1)\cosh |\zeta|-(e^{-i\omega t_1}-1)e^{i\theta}\sinh |\zeta|\big|^2\right]
    \nonumber\\
  &&~~~~
  +s_2
  \exp\left[-2{\lambda^2}
  \big|(e^{i\omega t_2}-1)\cosh |\zeta|-(e^{-i\omega t_2}-1)e^{i\theta}\sinh |\zeta|\big|^2\right]
  \nonumber\\
  &&~~~~
  +s_1s_2\cos\Theta(t_2,t_1)
\exp\left[-2\lambda^2\big|(e^{i\omega t_2}-e^{i\omega t_1})\cosh |\zeta|-(e^{-i\omega t_2}-e^{-i\omega t_1})e^{i\theta}\sinh |\zeta|\big|^2\right]\biggr).
\label{qssttsq}
\end{eqnarray}
When $\zeta$ takes a real number, Eq.~(\ref{qssttsq}) reduces to 
\begin{eqnarray}
  &&q_{s_1s_2}(t_1,t_2)={1\over 4}\biggl(1
  +s_1
  \exp\left[-8{\lambda^2}
  \sin^2{\omega t_1\over 2}(\cosh 2\zeta+\cos(\omega t_1)\sinh 2\zeta)
  \right]
    \nonumber\\
  &&~~~~
  +s_2
  \exp\left[-8{\lambda^2}
    \sin^2{\omega t_2\over 2}(\cosh 2\zeta+\cos(\omega t_2)\sinh 2\zeta)\right]
  \nonumber\\
  &&~~~~
  +s_1s_2\cos\Theta(t_2,t_1)
\exp\left[-8\lambda^2  \sin^2{\omega (t_2-t_1)\over 2}(\cosh 2\zeta+\cos(\omega (t_1+t_2))\sinh 2\zeta)\right]\biggr).
\label{qssttsq2}
\end{eqnarray}
Figures \ref{Squeeze} and \ref{Squeeze2} demonstrate the region where the two-time quasiprobability  (\ref{qssttsq2}) takes negative values on $t_1$ and $t_2$ planes, depending on a choice of $s_1$, $s_2$, and $\zeta$. 
The minimum value of the quasiprobability function (\ref{qssttsq2}) is 
approximately of the order
\begin{eqnarray}
  {\rm min} \{q_{s_1s_2}(t_1,t_2)\} \simeq -{\lambda^2\over 2} e^{2|\zeta|}.
\end{eqnarray}
In general, the squeezed initial condition boosts the signal of the Leggett-Garg inequalities violation, excepting the cases $s_1=s_2=1$ and $s_1=s_2=-1$ with $\zeta<0$.

\subsection{Connection with the experiment}
From Ref.~\cite{Carney}, we discuss the feasibility of signal detection. 
We introduced the mass density $\rho$ by $M=4\pi \rho\ell^3/3$ for the oscillator and
the approximation $L\sim \ell$, we have 
\begin{eqnarray}
&&\lambda^2={g^2\over \omega^2}={G^2m^2M\ell^2\over 2\omega^3\hbar \sqrt{L^2+\ell^2/4}^3}
\sim{G^2 m^2\rho\over\hbar \ell\omega^3},
\end{eqnarray}
which is estimated as
\begin{eqnarray}
{G^2 m^2\rho\over\hbar \ell\omega^3}
=1.7\times10^{-28}\biggl({m\over m_{\rm Cs}}\biggr)^2
\biggl({\rho\over 20 {\rm g/cm}^3}\biggr)
\biggl({\omega\over \omega_s}\biggr)^{-3}
\biggl({\ell\over 1{\rm mm}}\biggr)^{-1},
\end{eqnarray}
where $m_{\rm Cs}=2.2\times 10^{-25}$kg is the mass 
of a cesium atom and $\omega_s$ is defined as $\omega_s=2\pi/\tau$ 
with $\tau=10$ s. This was significantly a small signal, but
when we assumed the initial thermal state for the oscillator, 
the effective coupling constant was boosted by the factor  
$\bar n=k_BT/2\hbar \omega$ as
\begin{eqnarray}
\bar n\lambda^2
\sim0.5\times10^{-14}\biggl({m\over m_{\rm Cs}}\biggr)^2
\biggl({\rho\over 20 {\rm g/cm}^3}\biggr)
\biggl({\omega\over \omega_s}\biggr)^{-4}
\biggl({\ell\over 1{\rm mm}}\biggr)^{-1}
\biggl({T\over 300{\rm K}}\biggr).
\end{eqnarray}
The amplitude of the signal was the same as that discussed in Ref.~\cite{Carney}, 
in which the authors argued that 
the signal in the visibility function could be further
amplified by using many atoms and a coupling of the oscillator 
with another two-state system. 

For an experimental test of the violation of the Leggett-Garg inequalities, we need to measure the expectation values of $\langle \hat Q(t_j)\rangle={\rm Tr}[\bm n \cdot \bm \sigma (t_j) \rho_0]$
with $j=1,2$ and $C(t_2,t_1)={1\over 2}{\rm Tr}[\{\bm n\cdot \bm \sigma(t_1), \bm n\cdot \bm \sigma(t_2)\}\rho_0]$. The simplest case with $\bm n=(1,0,0)$ and $\Omega=0$ when we assumed the initial 
thermal state for the oscillator, we have
$\langle \hat Q(t_j)\rangle=e^{-8\lambda^2(2\bar n+1)\sin^2\omega (t_j/2)}$.
This expression is the same as the visibility function in Ref.~\cite{Carney}. 
Therefore, the measurement of $\langle \hat Q(t_j)\rangle$ is the same as that of 
the visibility function, which is essentially obtained by the two-state interference. 
On the other hand, $C(t_2,t_1)$ is the correlation function, which requires 
a much larger number of measurements to detect the signal with a sufficient statistical significance. 
This is a disadvantage of our approach with the Leggett-Garg inequalities for
testing the quantumness of gravity. 

However, as discussed in Refs.~\cite{Streltsov,Carney2,Maetal,Hosten}, the collapse and revival of the 
visibility function in an atomic interferometry could be generated by semi-classical models. 
The authors of Ref. \cite{Streltsov} demonstrated that an LOCC channel between a harmonic oscillator and a particle in a double well potential reproduces the collapse-and-revival dynamics in the interferometric signal.
Similarly, the authors of Ref. \cite{Maetal} demonstrated that the periodic collapses and revivals of the visibility can appear even when the oscillator is fully classical. Therefore, the revival of the 
visibility cannot be necessarily the signature of the quantumness of gravity connected to the entanglement. 
The Leggett-Garg inequality cannot be violated in a classical system, which will be a unique method  
to test a quantum property of gravity. It will be helpful that the signal of the violation of the Leggett-Garg
inequalities is boosted by preparing a squeezed initial state for the oscillator. 
The feasibility of detecting the signal against various noises is left for a future study. 

\section{Discussion}
We consider the Newton-Schr\"odinger approach in the present system to compare the 
difference of the predictions  in our theoretical model. 
In the Newton-Schr\"odinger approach, the gravitational potential $\Phi$ is given by 
expectation values of matter distributions with respect to the states.  
Explicitly, 
we may write the Newton-Schr\"odinger equation
\begin{eqnarray}
&&i{\partial |\psi(t)\rangle_{A}\over \partial t}
=\Bigl(\Omega \sigma^z+{GMm  \ell \over \sqrt{L^2+\ell^2/4}^3}\langle q\rangle \sigma^z\Bigr) |\psi(t)\rangle_{A},
\\
&&i{\partial |\psi(t)\rangle_{q}\over \partial t}
=\Bigl({p^2\over 2M}+{M\omega^2 \over 2}q^2+{GMm  \ell \over \sqrt{L^2+\ell^2/4}^3}\langle \sigma^z\rangle q\Bigr) |\psi(t)\rangle_{q},
\end{eqnarray}
for the state of the particle $|\psi(t)\rangle_{A}$ and the state of the oscillator
$|\psi(t)\rangle_{q}$, respectively, with which $\langle q\rangle$ and $\langle \sigma^z \rangle$ are defined by $\langle q\rangle={}_{q}\langle \psi(t)|q|\psi(t)\rangle_{q}$ 
and $\langle \sigma^z \rangle={}_{A}\langle \psi(t)|\sigma^z|\psi(t)\rangle_{A}$, respectively.
Here $p$ is the conjugate momentum of $q$.

For the initial state of the oscillator and the particle adopted in our analysis (for example, the initial state given by \eqref{isvacuum}), the gravitational interaction vanishes, i.e., 
$\langle q\rangle=\langle \sigma^z \rangle=0$, because of the symmetry of the system. 
When the Larmor precession-like frequency vanishes,
$\Omega=0$, there are no violation of the Leggett-Garg inequalities in the Newton-Schr\"odinger approach. 
The violation of the Leggett-Garg inequalities which appears via the gravitational interaction 
in the previous section can be regarded as a consequence of the quantum nature of the gravitational
interaction.

\section{Conclusions}
We investigated the violation of the Leggett-Garg inequalities due to the gravitational interaction in the hybrid system \cite{Carney} using a two-time quasiprobability. 
With the initial time $t_1=0$, we first discussed the role of the gravitational interaction in the violation of the Leggett-Garg inequalities of the two-time quasiprobability in the connection to the entanglement generated by the gravitational interaction.
In the case $\Omega\neq0$, the Larmor precession-like behavior appears, and we can 
assume the parameters so that the Leggett-Garg inequalities are violated when the gravitational interaction is switched off.
This violation of the Leggett-Garg inequalities is due to the quantum property 
of the particle system itself. In this setup, $t_1=0$ and $\Omega\neq0$, we 
demonstrated that the entanglement,
induced by the gravitational interaction switched on, 
suppresses the violation of the Leggett-Garg inequalities. Furthermore, in some parameter settings, the quasiprobability equals the square of the entanglement negativity. 

When the Larmor precession-like behavior in the two spatially localized 
states was switched off,  i.e., $\Omega=0$, we demonstrated that the quasiprobability 
took negative values due to the gravitational interaction, in general, 
depending on the choice of the parameters 
and the initial conditions. For the realistic situation $g/\omega\ll1$, 
the minimum value of the two-time quasiprobability was 
of the order $-g^2/2\omega^2$ when the initial state of 
the oscillator was in the ground state, while it was 
of the order $-g^2\bar n/\omega^2$ when the initial 
state of  the oscillator was in the thermal state, 
where $\bar n=k_BT/2\omega$.
As discussed in Ref.~\cite{Carney}, the choice of the initial thermal state significantly increases the signal of the quasiprobability owing to gravitational interaction. We also demonstrated that squeezing the
initial state of the oscillator significantly boosts the 
amplitude of the signal of the Leggett-Garg inequalities violation. 

Here, we discuss the origin of the violation of the Leggett-Garg inequalities due to gravity in the hybrid system that was determined in Sec. III B, C, and D. 
The violation of the Leggett-Garg inequalities in the case where $\Omega=0$ originates from the gravitational interaction;
otherwise, no evolution arises in the system of the particle. Gravitational interaction generates an entangled hybrid cat state 
Eq.~(\ref{ehqsc}), therefore, entanglement plays an important role in the Leggett-Garg inequalities violation. 
In the Leggett-Garg inequalities violation, the terms $\langle Q(t_1) \rangle$ and $\langle Q(t_2) \rangle$ play
a crucial role in making the two-time quasiprobability negative values. For the simplest case, $\bm n=(1,0,0)$, we have
$\langle Q(t) \rangle={\rm Tr}[\bm n\cdot\sigma(t)\rho_0 ]=e^{-8\lambda^2\sin^2(\omega t/2)}$, which is a visibility function 
addressed in the Ref. \cite{Carney}.  Based on Ref.~\cite{Carney,Carney2}, the oscillatory behavior of the
visibility function originates from non-separable 
evolution of the state owing to the gravitational interaction, 
which causes the entanglement of the system.
For the case $\Omega=0$, the Leggett-Garg inequalities are not violated when the oscillator and the particle undergo 
the separable unitary evolution
with the separable initial state.
Therefore, it can be concluded that the Leggett-Garg inequalities violation for the case $\Omega=0$ is derived from the non-separable property of the gravitational interaction. 

However, the origin of the violation of the Leggett-Garg inequalities may still remain
a room for discussions. For the case $t_1=0$ and $\Omega\neq0$, the gravitational interaction causes the entanglement, which always suppresses the violation of the Leggett-Garg inequalities caused by 
the quantum nature of the particle system itself.
For the case $\Omega=0$, the gravitational interaction only causes the evolution 
in the particle system, which causes the entanglement between the particle and the oscillator
as longs as $\lambda\neq0$.
Therefore, we concluded that the origin of the violation of the Leggett-Garg inequalities 
is the gravitational interaction and the entanglement induced by the gravitational interaction.
This is supported by the result of Sec.~IV that the Newton-Schr\"dinger approach does not 
cause the violation of the Leggett-Garg inequalities in which the gravitational interaction causes no entanglement.
However, the gravitational entanglement has two effects, i.e., violation and holding of the Leggett-Garg inequalities depending on the parameter $t_1$ and $t_2$. This can be understood 
from Eq.~(\ref{demo2}). Namely, the two-time quasiprobability is expressed by the latter term
in proportion to $\lambda^2$ in Eq.~(\ref{demo2}) when $s_1$ and $s_2$ are adopted as those in the panels of Fig.~\ref{Gmacroc}. When the two-time quasiprobability takes negative/positive values, the 
Leggett-Garg inequalities are violated/satisfied. 
We haven't clarified how these different aspects of the entanglement due to the gravitational interaction appears in the violation/holding of the Leggett-Garg inequalities in an intuitive manner.
Furthermore, the particle is equipped with quantum properties. Therefore, it might 
be difficult to exclude the possibility that the violation of the Leggett-Garg inequalities 
comes from the quantumness of the particle system itself. 

In general, it is interesting to test quantum properties of macroscopic systems
to know the boundary between quantum systems and classical systems. 
Our research, which is motivated by testing quantum properties of the gravitational 
interaction, can be regarded as a test of the quantum aspects of a gravitational 
potential as a macroscopic system through the Leggett-Garg inequalities.
The Leggett-Garg inequalities are originally developed on the basis of the macroscopic 
realism and the noninvasive measurability, which are tested by a measurement of the violation of the inequalities. 
In our system, a superposition state of the macroscopic oscillator is generated 
by the superposition state of the particle initially prepared. When the initial state of 
the oscillator is prepared as a superposition state of coherent states by some method, e.g., $(|\xi_0\rangle_C+|\xi_1\rangle_C)/\sqrt{2}$ with coherent parameters $\xi_0$ and $\xi_1$, an entangled state between the oscillator and the particle will appear, as is shown in the Appendix~A. 
The result Eq.~(\ref{EqSecV}) means that the particle system could be used as a probe of the superposition state of the oscillator by measuring an interference of the particle state caused by the entanglement. 
When the particle and the oscillator interact through 
a different force, the factor will be written in a corresponding form reflecting the
different interaction. Therefore, a particle in a superposition state could be 
a probe of a quantum state of the macroscopic oscillator and the quantum nature of the
interaction when the interaction between them is well understood. 
It is interesting to investigate the violation of the Leggett-Garg inequalities in the particle's state as a probe of quantum aspects of macroscopic oscillators and their interaction, which is left as future investigations.

\acknowledgments
We thank S. Maeda, Y. Kaku, and Y. Osawa for useful discussions.
We also thank D. Miki and S. Iso for useful discussion and communication, which significantly improved the manuscript. 
Y.N. was partially supported by JSPS KAKENHI, Grant No. 19K03866. 

\appendix
\section{Result with other initial state for oscillator}
When the initial state of the system is prepared as 
\begin{eqnarray}
|\psi_0\rangle={1\over 2}\bigl(|0\rangle_A+|1\rangle_A\bigr)\otimes\bigl(|\xi_0\rangle_C+|\xi_1\rangle_C\bigr),
\label{isvacuumA}
\end{eqnarray}
where $|\xi_j\rangle_C$ for $j=0,1$ is a coherent state of the oscillator, 
the state will evolve as\begin{eqnarray}
|\psi(t)\rangle =U(t)|\psi_0\rangle=
e^{-i(\Omega\sigma^z+\omega  a^\dagger a)t} e^{g\sigma^z(\alpha(t)a-\alpha^*(t)a^\dagger)}|\psi_0\rangle.
\end{eqnarray}
Using the formula, 
$
D(-g\sigma^z\alpha^*(t))D(\xi_j)=e^{ g\sigma^z(-\alpha^*(t)\xi_j^*+\alpha(t)\xi_j)/2}
D(-g\sigma^z\alpha^*(t)+\xi_j)$, 
we have
\begin{eqnarray}
e^{-i\omega a^\dagger a}D(-g\sigma^z\alpha^*(t))D(\xi_j)|0\rangle&=&e^{ g\sigma^z(-\alpha^*(t)\xi_j^*+\alpha(t)\xi_j)/2}
e^{-i\omega a^\dagger a}D(-g\sigma^z\alpha^*(t)+\xi_j)|0\rangle
\nonumber
\\
&=&e^{g\sigma^z(-\alpha^*(t)\xi_j^*+\alpha(t)\xi_j)/2}
|g\sigma^z\alpha(t)+ \xi_j e^{-i\omega t}\rangle_C,
  \label{UnitaryA3}
\end{eqnarray}
which leads to 
\begin{eqnarray}
&&|\psi(t)\rangle ={1\over 2}\Bigl(
e^{ g(-\alpha^*(t)\xi_0^*+\alpha(t)\xi_0)/2}
|0\rangle_A
|g\alpha(t)+ \xi_0 e^{-i\omega t}\rangle_C
+
e^{-g(-\alpha^*(t)\xi_0^*+\alpha(t)\xi_0)/2}
|1\rangle_A
|-g\alpha(t)+ \xi_0 e^{-i\omega t}\rangle_C
\nonumber\\
&&~~~~~~~~~~~
+
e^{ g(-\alpha^*(t)\xi_1^*+\alpha(t)\xi_1)/2}
|0\rangle_A
|g\alpha(t)+ \xi_1 e^{-i\omega t}\rangle_C
+
e^{-g(-\alpha^*(t)\xi_1^*+\alpha(t)\xi_1)/2}
|1\rangle_A
|-g\alpha(t)+ \xi_1 e^{-i\omega t}\rangle_C\Bigr).
\end{eqnarray}
When $g/\omega\ll |\xi_j|$ for $j=0,1$, the state can be 
approximately written as
\begin{eqnarray}
&&|\psi(t)\rangle \simeq{1\over 2}\bigl(
e^{ g(-\alpha^*(t)\xi_0^*+\alpha(t)\xi_0)/2}
|0\rangle_A
|\xi_0 e^{-i\omega t}\rangle_C
+
e^{-g(-\alpha^*(t)\xi_0^*+\alpha(t)\xi_0)/2}
|1\rangle_A
| \xi_0 e^{-i\omega t}\rangle_C
\nonumber\\
&&~~~~~~~~~~~~
+
e^{ g(-\alpha^*(t)\xi_1^*+\alpha(t)\xi_1)/2}
|0\rangle_A
|\xi_1 e^{-i\omega t}\rangle_C
+
e^{-g(-\alpha^*(t)\xi_1^*+\alpha(t)\xi_1)/2}
|1\rangle_A
| \xi_1 e^{-i\omega t}\rangle_C\bigr).
\label{EqSecV}
\end{eqnarray}
We note that the result is an entangled state between the oscillator 
and the particle with the factor $e^{\pm g(-\alpha^*(t)\xi_j^*+\alpha(t)\xi_j)/2}$ with $j=0,1$, which
comes from the gravitational interaction between them.

\end{document}